\documentclass[12pt,a4paper,onecolumn]{iopart}
\usepackage{graphics,graphicx}
\usepackage[small]{caption}
\usepackage{amssymb}
\usepackage[latin1]{inputenc}

\newcommand{\LaF}{LaO$_{0.94}$F$_{0.06}$FeAs}

\begin{document}
\title{Electromagnetic response of \LaF:\\ \emph{AC} susceptibility and microwave surface resistance}

\author{A Agliolo Gallitto$^1$, G Bonsignore$^1$, M Bonura$^1$, M Li Vigni$^1$,\\ J  L Luo$^2$ and A F Shevchun$^3$}

\address{$^1$CNISM and Dipartimento di Scienze Fisiche ed Astronomiche, Universit\'{a} di Palermo,\\ Via Archirafi 36, I-90123 Palermo, Italy}

\address{$^2$Beijing National Laboratory for Condensed Matter Physics, Institute of Physics, Chinese Academy of Sciences, Beijing 100190, China}

\address{$^3$Institute of Solid State Physics, Russian Academy of Sciences, Chernogolovka, Moscow District}

\ead{gaetano.bonsignore@fisica.unipa.it}

\begin{abstract}
We discuss on the electromagnetic response of a polycrystalline sample of \LaF\ exposed to \emph{DC} magnetic fields up to $10$~kOe. The low- and high-frequency responses have been investigated by measuring the \emph{AC} susceptibility at 100~kHz and the microwave surface resistance at 9.6~GHz. At low as well as high \emph{DC} magnetic fields, the susceptibility strongly depends on the amplitude of the \emph{AC} driving field, highlighting enhanced nonlinear effects. The field dependence of the \emph{AC} susceptibility exhibits a magnetic hysteresis that can be justified considering the intragrain-field-penetration effects on the intergrain critical current density. The microwave surface resistance exhibits a clockwise magnetic hysteresis, which cannot be justified in the framework of the critical-state models of the Abrikosov-fluxon lattice; it may have the same origin as that detected in the susceptibility.
\end{abstract}

\section{Introduction}
It is by now accepted that polycrystalline superconductors (\emph{SC}) can be modeled as superconducting grains connected by weak links~\cite{gomory,Saha,SUST_93}. The system of intergrain contacts forms an effective pinning medium, generally indicated as intergrain region. The weak coupling of the intergrain region strongly affects the electromagnetic (\emph{em}) response, especially when a \emph{DC} magnetic field is superimposed to the \emph{AC} driving field. Weak magnetic fields, even smaller than the lower critical field of grains, $H_{c1}$, easily penetrates the intergrain region through Josephson vortices. On increasing the field above $H_{c1}$, flux penetrates the superconducting grains and further  mechanisms, of flux penetration and energy dissipation, come into play.

The granularity effects in the \textit{in-field} \emph{em }response of ceramic cuprate \emph{SC} have been investigated and discussed by several authors~\cite{gomory,Saha,SUST_93,Evetts,Muller_PhysC168,Muller_isteresi}. Soon after the discovery of superconductivity in iron-based pnictides, some authors have highlighted \emph{em}-granularity effects~\cite{polichetti,granularityAPL}. In this paper, we report a study of the in-field \emph{em }response of a polycrystalline sample of \LaF. We have measured the \emph{AC} susceptibility, $\chi =\chi^{\prime}+i \chi^{\prime \prime}$, at 100~kHz and the microwave (\emph{mw}) surface resistance, $R_s$, at 9.6~GHz. In order to obtain information on the intergrain and/or intragrain vortex dynamics, $\chi$ and $R_s$ has been investigated as a function of the temperature, at fixed applied fields, and as a function of the magnetic field, at fixed temperatures. At low as well as high \emph{DC} magnetic fields, $\chi$ strongly depends on the amplitude of the \emph{AC} field, highlighting enhanced nonlinear effects. The field dependence of $\chi$ exhibits a magnetic hysteresis that can be justified considering the intragrain-field-penetration effects on the intergrain critical current density. The field dependence of $R_s$ exhibits a clockwise hysteresis similar to that observed in $\chi^{\prime}$; we suggest that the hysteretic behaviors of $\chi$ and $R_s$ have the same origin and, in particular, are due to the granular nature of the sample.

\section{Experimental and Sample}\label{apparatus}
The \emph{em }response has been measured in a polycrystalline sample of \LaF. The  specimen was prepared by solid state reaction using LaAs, Fe$_2$O$_3$, Fe and LaF$_3$ as starting materials; details on the preparation and properties of the sample are reported in Ref.~\cite{luo}.

The \emph{AC} susceptibility at 100 kHz has been measured by a standard two-coil susceptometer~\cite{Nikolo_base}, coupled to an \emph{hp}-4263 B LCR meter. The sample is located at the center of one of the secondary coils by a sapphire holder, at which a temperature sensor and a heater are fixed. The \emph{mw} surface resistance has been measured by the cavity-perturbation technique~\cite{noiisteresi}. A cylindrical copper cavity is tuned in the TE$_{011}$ mode~\cite{lanc} resonating at 9.6 GHz; the sample is located at the center of the cavity, where the \emph{mw} magnetic field is maximum.

The susceptometer (or the cavity) is placed between the poles of an electromagnet which generates \emph{DC} magnetic fields up to $H_0=10$~kOe. Two additional coils, independently fed, allow compensating the residual field, within 0.01~Oe, and working at low magnetic fields.

\section{Results and Discussion}
\subsection {AC Susceptibility}\label{SUSC}
The real and imaginary components of the \emph{AC} susceptibility at 100~kHz have been investigated as a function of the temperature (from $T=4.2$~K up to $T\approx 30$~K), the amplitude of the \emph{AC} field (up to $H_{ac} = 3$~Oe) and the \emph{DC} magnetic field (at increasing and decreasing values), in the field geometry $\emph{\textbf{H}}_{0}\parallel\emph{\textbf{H}}_{ac}$.
\begin{figure}[b]
\begin{center}
\includegraphics[width=0.8\textwidth]{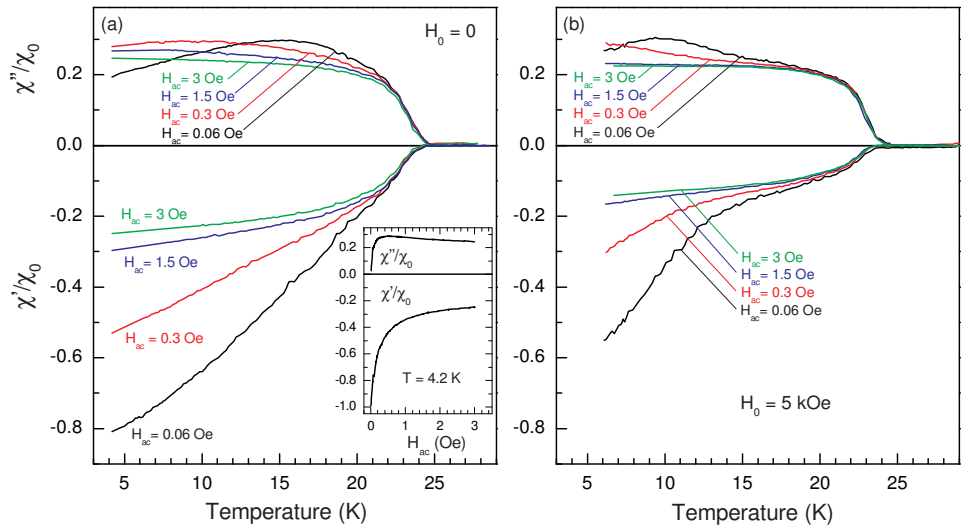}
\end{center}\vspace{-0.5cm}
\caption{\label{CHI(T)}Real and imaginary components of the \emph{AC} susceptibility at 100~kHz as a function of the temperature, obtained at different value of $H_{ac}$. (a) $H_0=0$; (b) $H_0=5$~kOe. Inset:  $\chi^{\prime}$ and $\chi^{\prime \prime}$ as a function of $H_{ac}$, at $H_0=0$ and $T=4.2$~K. }\vspace{-5mm}
\end{figure}
Fig.~\ref{CHI(T)} shows the temperature dependence of $\chi^{\prime}$ and $\chi^{\prime \prime}$, obtained in zero-field-cooled (\emph{ZFC}) sample, at $H_0=0$ (a) and $H_0=5$~kOe (b) for different values of $H_{ac}$. The inset shows $\chi^{\prime}$ and $\chi^{\prime \prime}$ as a function of $H_{ac}$, at $H_0=0$ and $T=4.2$~K. All the curves are normalized to the value $\chi_0$ obtained extrapolating the curves $|\chi^{\prime}(H_{ac})|$ to $H_{ac}\rightarrow 0$, at $T=4.2$~K and $H_0=0$.

From Fig.~\ref{CHI(T)}, whether for $H_0=0$ or for $H_0=5$~kOe, one can note a near-$T_c$ region characterizing the linear response ($\chi$ independent of $H_{ac}$) and a region at lower temperatures highlighting enhanced nonlinear effects. For $H_0=0$, the linear response is ascribable to reversible flux penetration in the superconducting grains; the nonlinear response is due to intergrain critical state~\cite{gomory,Muller_PhysC168,Lee}. The peak in $\chi^{\prime \prime}$ corresponds to the maximum energy losses, occurring when $H_{ac}$ penetrates to the center of the sample through the intergranular region. As expected, on increasing $H_{ac}$ the loss peak moves toward lower temperatures; however, the peak shift is much more enhanced than that observed in cuprate \emph{SC}~\cite{Muller_PhysC168,Lee}.

From $\chi^{\prime \prime}(H_{ac})$ of Fig.~\ref{CHI(T)} (a), the intergrain full penetration field, $H_m^*$, at $T=4.2$~K results about 0.6~Oe. This value is very small, suggesting a extremely low intergrain critical current density, $J_{cm}$. However, we would like to remark that in this sample we have detected a strong frequency dependence of the $\chi^{\prime \prime}(T)$-peak position~\cite{noiChi(T)}. In particular, at $10$~kHz, $H_m^*(4.2~\mathrm{K})\approx 0.1$~Oe. This frequency dependence indicates relaxation of the intergrain critical state with characteristic times lower than $10^{-5}$~s, and leads to underestimate $J_{cm}$.

At $H_0>H_{c1}$, Abrikosov fluxons penetrate the grains affecting the $\chi(T)$ features. In this case, two peaks in $\chi^{\prime \prime}(T)$, and two steps in $\chi^{\prime}(T)$, are expected. The low-$T$ peak is due to full penetration of $H_{ac}$ in the intergrain region and the high-$T$ peak is due to full penetration of $H_{ac}$ in the grains~\cite{gomory,Lee}. We have performed measurements up to $H_0=10$~kOe, but two well distinct peaks have never been observed; this could be caused by a wide grain-size distribution, which makes the high-$T$ peak broadened and embedded in the tail of the intergrain $\chi^{\prime \prime}(T)$ peak. Therefore, the results of Fig.~\ref{CHI(T)} (b) may be justified assuming that the peak at $T\approx 9$~K is due to the intergrain full penetration of $H_{ac}$ and the \textit{kink} at $\sim 20$~K to the intragrain full penetration.

\begin{figure}[b!]
\begin{center}
\includegraphics[width=0.85\textwidth]{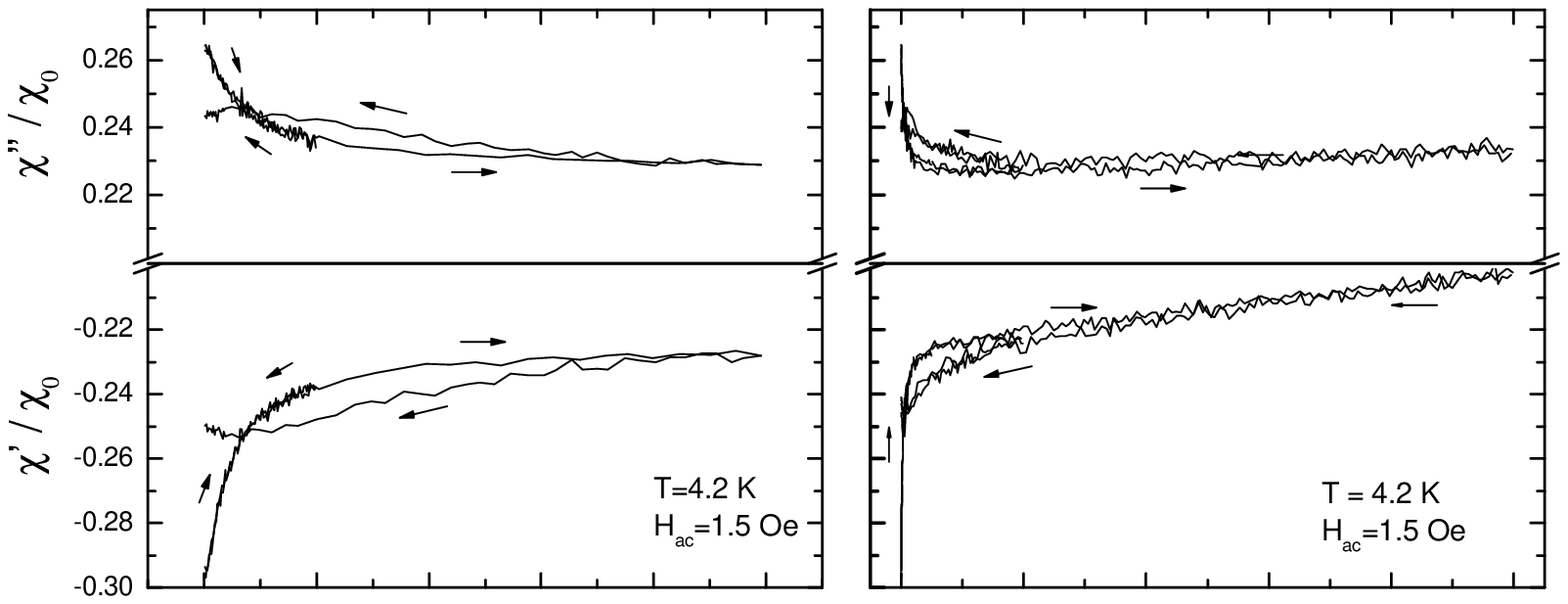}\vspace{-2mm}
\includegraphics[width=0.855\textwidth]{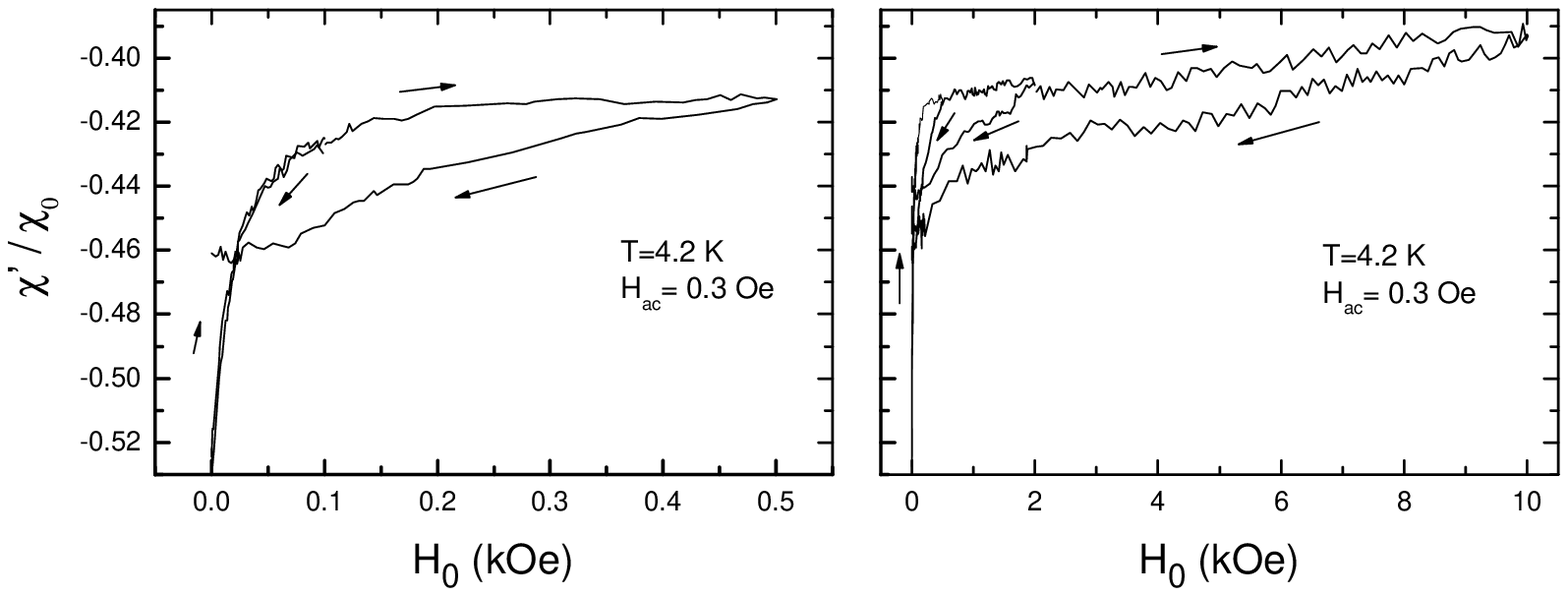}
\end{center}\vspace{-0.5cm}
\caption{\label{isteresi}$\chi^{\prime}$ and $\chi^{\prime \prime}$ as a function of the \emph{DC} magnetic field, obtained sweeping $H_0$ from 0 to $H_{max}$ and back, for different values of $H_{max}$.}\vspace{-5mm}
\end{figure}

In Fig.~\ref{isteresi}, we report the \emph{DC} field dependence of the \emph{AC} susceptibility, obtained at $T=4.2$~K by sweeping $H_0$ from 0 to $H_{max}$ and back, for different values of $H_{max}$. All the curves show a reversible behavior for $H_{max}\leq 100$~Oe; when $H_{max}$ slightly overcomes 100~Oe, a hysteresis appears. The hysteresis-loop amplitude depends on $H_{ac}$. As one can see, for $H_{ac}=1.5$~Oe no hysteresis is observed at $H_0 \gtrsim 5$~kOe; on the contrary, for $H_{ac}=0.3$~Oe the hysteresis in $\chi ^{\prime}$ is noticeable in the whole range of \emph{DC} fields investigated. We do not report the field dependence of $\chi^{\prime \prime}$ obtained at $H_{ac}=0.3$~Oe because the signal is very weak and the curve is very noisy; however, we would like to remark that at low \emph{AC} fields the hysteresis in $\chi^{\prime \prime}$ reduces. It is worth noting that the increasing-field branch of the $\chi^{\prime \prime}(H_0)$ curve obtained at $H_{ac}=1.5$~Oe exhibits a monotonic decrease, this occurs because $H_{ac}>H_m^*$. Measurements performed at $H_{ac}<H_m^*$ have shown an initial rise and a following  decrease, as expected.

The hysteretic behavior of $\chi(H_0)$ can be explained by the models discussed  in~\cite{Saha,Evetts,Muller_isteresi}, in the framework of which the hysteresis in $\chi$ is related to the hysteresis of $J_{cm}$. The magnetic history of $J_{cm}$ can be qualitatively described as follows. On increasing $H_0$ from the \emph{ZFC} condition, the current through the intergrain region decreases according to the Josephson-junction behavior; a further reduction occurs when $H_0$ reaches the first-penetration field of grains, because Abrikosov-fluxon gradient inside the grains produces a diamagnetic moment. On decreasing $H_0$ from $H_{max}$, the moment of grains becomes paramagnetic and reduces the average effective field at the junctions; this makes $J_{cm}(H_0)$ larger. When $H_0$ reaches the value of the paramagnetic contribution, the decreasing-field branch of the $J_{cm}(H_0)$ curve shows a peak, whose position depends on $H_{max}$ and shifts toward higher fields when $H_{max}$ increases. According to these considerations and looking at Fig.~\ref{isteresi}, we infer that the first-penetration field of grains is $\approx 100$~Oe ($\chi(H_0)$ is reversible for $H_{max} \leq 100$~Oe). Moreover, the clockwise hysteresis of $\chi^{\prime}(H_0)$ and the counterclockwise one of $\chi^{\prime \prime}(H_0)$, as well as their dependence on $H_{ac}$, can be justified in the framework of the cited models.

We have investigated the field dependence of $\chi$ at different temperatures, from 4.2~K up to $T_c$. On increasing the temperature, the amplitude of the hysteresis loop decreases and the range of $H_0$ in which it is detectable shrinks; eventually, at $T\approx 20$~K the $\chi(H_0)$ curve exhibits a reversible behavior in the whole range of fields investigated. This result is consistent with those of Fig.~\ref{CHI(T)}, in which a linear response is detected at $T\approx 20$~K, which is ascribable to the intragrain contribution.

\subsection{Microwave Surface Resistance}\label{Rs}
The \emph{mw} surface resistance, $R_s$, has been measured in the \emph{ZFC} \LaF\ sample as a function of the temperature, at fixed $H_0$ values, and as a function of $H_0$ at fixed temperatures, in the field geometry $\emph{\textbf{H}}_{0}\perp \emph{\textbf{H}}_{ac}$. The measurements have been performed at very low input power; the estimated amplitude of the \emph{mw} magnetic field is of the order of 1~mOe.
\begin{figure}[h]
\begin{minipage}{18pc}\centering
\includegraphics[width=16pc]{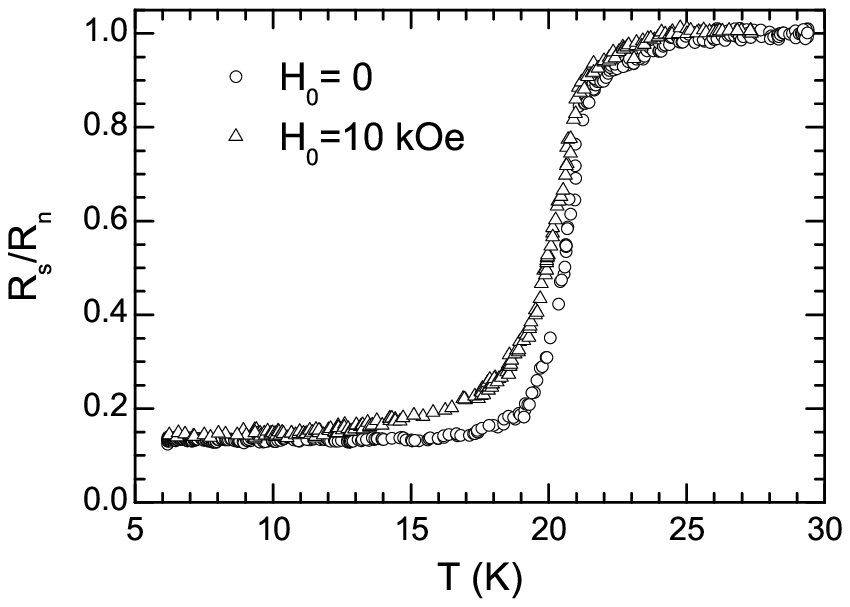}
\caption{\label{Rs(T)}$R_s/R_n$ as a function of temperature in the \emph{ZFC} sample, at $H_0=0$ and $H_0 =10$~kOe.}
\end{minipage}\hspace{2pc}%
\begin{minipage}{18pc}\centering
\includegraphics[width=16pc]{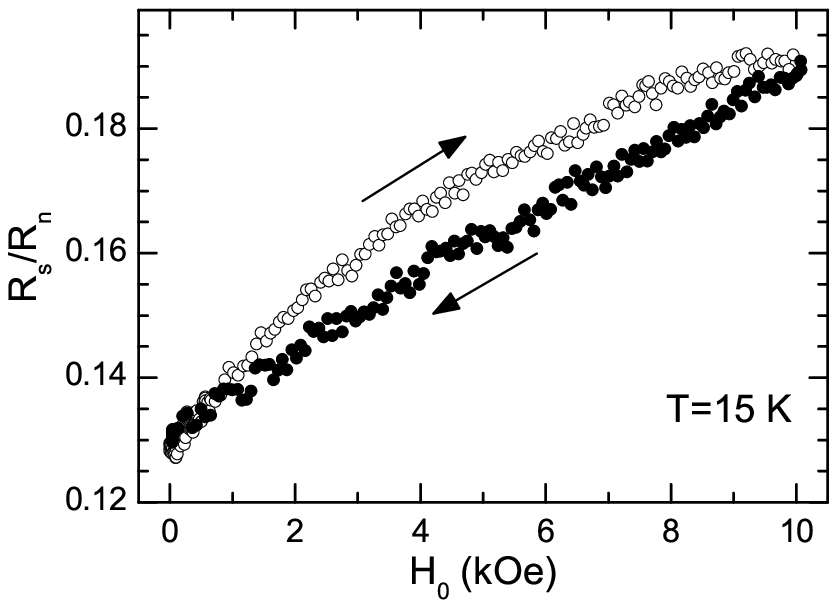}
\caption{\label{Rs(H)}$R_s/R_n$ as a function of $H_0$, obtained by sweeping the field up to 10~kOe and back.}
\end{minipage}\vspace{-2mm}
\end{figure}

Fig.~\ref{Rs(T)} shows the temperature dependence of $R_s/R_n$ obtained at $H_0=0$ and $H_0=10$~kOe; $R_n$ is the \emph{mw} surface resistance at $T\gtrsim T_c$. Considering the geometry factor of the sample, we estimate $R_n\approx 0.5~\Omega$ and the residual \emph{mw} surface resistance, at $T=4.2$~K and $H_0=0$, $R_{res}\approx 65~\mathrm{m}\Omega$; the value of $R_{res}$ is very high and confirms that a large number of weak links contribute to the \emph{mw} energy dissipation. As one can see, the variation of $R_s$ induced by a magnetic field of 10~kOe is very low up to temperatures $\lesssim T_c/2$ most likely because of the high upper-critical-field value of these \emph{SC}; this is consistent with the not-visible shift of $T_c$ with $H_0$.

Fig.~\ref{Rs(H)} shows the normalized field-induced variations of $R_s$ of the \emph{ZFC} sample, obtained at $T=15$~K by sweeping $H_0$ from 0 up to 10~kOe and back. By analyzing the experimental data at low $H_0$ values, one notes that $R_s$ starts to vary at $H_0$ lower than the expected value of the first-penetration field of Abrikosov fluxons. This behavior is consistent with the results of the \emph{AC} susceptibility and shows that also at \emph{mw} frequency the presence of weak links significantly contributes to the energy losses. Moreover, one can see a clockwise magnetic hysteresis similarly to what observed in the $\chi^{\prime}(H_0)$ curve.

Magnetic hysteresis in the \emph{mw} surface resistance of type-II \emph{SC} in the mixed state has been detected by different authors~\cite{noiisteresi,sridhar,Ji}; it has been ascribed to the different value of the magnetic induction inside the sample at increasing and decreasing \emph{DC} fields, due to the critical state of the fluxon lattice. In Refs.~\cite{noiisteresi,noistatocritico}, we have investigated the effect of the critical state of the Abrikosov-fluxon lattice in $R_s(H_0)$ and have quantitatively described the hysteretic behavior of $R_s(H_0)$. Since the field-induced variations of $R_s$ are related to the presence and motion of fluxons, it is expected that the shape of the hysteresis is strongly related to that of the magnetization curve, giving rise to a higher value of $R_s$ in the decreasing-field branch with respect to that of the increasing-field branch. So, taking into account the effects of the critical state of the Abrikosov-fluxon lattice, a clockwise hysteresis cannot be explained. Another result of our investigation is that the amplitude of the hysteresis is proportional to the full penetration field, $H^*$: samples of small size and/or small $J_c$ are expected to exhibit weak hysteretic behavior.

Clockwise hysteresis in $R_s(H_0)$ has been detected in granular cuprate \emph{SC} and has been quantitatively discussed by Ji et \textit{al.}~\cite{Ji} in terms of the so-called two-level critical-state model, which considers a large critical current density inside the grains, $J_{cg}$, and a much weaker $J_{cm}$. The authors calculate the intergrain fluxon density and show that the contribution to the \emph{mw} losses of the intergrain fluxons accounts for the clockwise hysteresis of $R_s(H_0)$.

We would like to remark that the model elaborated by Ji et \textit{al.} for explaining the feature of the $R_s(H_0)$ curve and those~\cite{Evetts,Muller_isteresi} we have discussed in Sec.~\ref{SUSC} to account for the hysteretic behavior of $\chi(H_0)$, assume that the main contribution responsible for the in-field \emph{em }response comes from the intergrain fluxon dynamics. In principle, in order to calculate the field-induced variation of $R_s$ one should take into account two contributions; one arising from critical state of intragrain fluxons, which should give rise to a counterclockwise hysteresis, and another arising from critical state of intergrain fluxons, which should give rise to a clockwise hysteresis. Our results suggest that for \emph{DC} fields up to 10~kOe the intragrain contribution in this sample is negligible. Two different reasons may be responsible of the negligible intragrain contribution to the hysteresis of $R_s(H_0)$: i) the applied magnetic field is much lower than $H_{c2}$ with consequent low fluxon density in the grains; ii) the grains have small $H^*$ because of their small size with consequent small trapped flux.

\section{Conclusion}
We have investigated the low and high frequency response of a \LaF\ sample exposed to \emph{DC} magnetic fields up to 10~kOe, by measuring the \emph{AC} susceptibility at 100~kHz and the microwave surface resistance at 9.6~GHz. The results have been qualitatively discussed in the framework of models reported in the literature. We have shown that the \emph{em }response is strongly affected by the granular nature of the sample. Both at zero and at high \emph{DC} magnetic fields, we have highlighted nonlinear effects strictly related to the critical state of fluxons in the intergrain region.

Measurements of the susceptibility at zero \emph{DC} magnetic field as a function of the \emph{AC} driving field have highlighted values of the intergrain full-penetration field smaller than 1~Oe, even at the lowest temperature investigated. This very small value of $H_m^*$, although further reduced by magnetic-relaxation during the \emph{AC} cycle, indicates extremely low pinning potential. At \emph{DC} magnetic fields higher than 100~Oe, both $\chi^{\prime}$ and $\chi^{\prime \prime}$ show a magnetic hysteresis, which can be justified considering the intragrain-field-penetration effects on the intergrain critical current density.

The field dependence of the microwave surface resistance exhibits a clockwise hysteresis, similar to that detected for $\chi^{\prime}$, which cannot be justified in the framework of critical-state models of the Abrikosov-fluxon lattice. According to the models reported in the literature for granular superconductors, the clockwise magnetic hysteresis of $R_s$ can be accounted for by supposing that the main contribution to the field-induced \emph{mw} losses comes from the intergrain region.

\ack{Work partially supported by the University of Palermo in the framework of the International Co-operation Project CoRI 2007 Cupane.}

\end{document}